\documentclass{jpsj-suppl}
\usepackage{txfonts} 

\title{Flow-History-Dependent Behavior 
in Entangled Polymer Melt Flow with Multiscale Simulation}

\author{Takahiro \textsc{Murashima}$^{1,2}$ 
and Takashi \textsc{Taniguchi}$^{1,2}$}

\inst{$^{1}$Department of Chemical Engineering, Kyoto University,
Katsura Campus, Nishikyo-ku, Kyoto 615-8510, Japan \\
$^{2}$Core Research for Evolutional Science and Technology, JST, 
Kawaguchi, Saitama 332-0012, Japan}

\email{murasima@cheme.kyoto-u.ac.jp}

\recdate{October 3, 2011}

\abst{Polymer melts represent the flow-history-dependent behavior.
To clearly show this behavior,
we have investigated flow behavior of an entangled polymer melt 
around two cylinders placed in tandem along the flow direction 
in a two dimensional periodic system.
In this system, the polymer states around a cylinder 
in downstream side are different from the ones around another cylinder
in upstream side because the former ones have 
a memory of a strain experienced when passing around the cylinder in
upstream side but the latter ones do not have the memory.
Therefore, the shear stress distributions around two cylinders
are found to be different from each other.
Moreover, we have found that the averaged flow velocity 
decreases accordingly with increasing the distance between two cylinders
while the applied external force is constant. 
While this behavior is consistent with that of the Newtonian fluid,
the flow-history-dependent behavior enhances 
the reduction of the flow resistance.}

\kword{multiscale simulation, fluid particle simulation, coarse-grained
simulation, memory effect}

\begin{document}
\maketitle

\section{Introduction}
Predicting flow of
entangled polymer melts is difficult
because the microscopic states of polymers depend 
on the flow history, namely 
the history of the experienced strain during the flow.
Moreover, 
the quite large number of the degrees of freedom in 
the entangled polymer melt makes it difficult
to build the constitutive equation 
that describes the relation among 
the stress tensor and the strain (or strain-rate) tensor
in the entangled polymer melt.
To overcome the difficulty existing 
in the flow prediction of entangled polymer melts,
we have developed a new multiscale simulation~\cite{MT2011}
composed of the macroscopic fluid particle simulation~\cite{MSPH,FPM}
and the microscopic entangled polymer dynamics 
simulation~\cite{PCNM,DSLM_PRL,DSLM,SSSM,SSSM_GPU}.
Using the fluid particle simulation 
where each fluid particle has 
a polymer simulator that describes the polymer states 
in the fluid particle itself,
we can accurately consider the flow history of 
the fluid particle and the polymers in the fluid particle.
The entangled polymer dynamics
simulations~\cite{PCNM,DSLM_PRL,DSLM,SSSM,SSSM_GPU} 
are based on 
reptation
theory~\cite{ReptationTheory1,ReptationTheory2,ReptationTheory3}
where the dynamics of a polymer chain is constrained 
in a tube created by the surrounding polymers
because of excluded volume effect and entanglements between polymers.
Because each polymer chain in the entangled polymer dynamics simulation
is described with the number of 
hypothetical entanglement points $Z$ and 
the tube segments $\boldsymbol{r}^{\rm s}_j (j=1,\cdots,Z)$
connecting the entanglement points on the chain,
we can decrease considerably large amount of the degrees of freedom
into a manageable number of the degrees of freedom.
The multiscale simulation enables us to simulate 
the polymer melt flow with remaining 
the detailed information on the entangled states
of polymers.

\begin{figure}[t]
\begin{center}
\includegraphics[width=0.5\columnwidth]{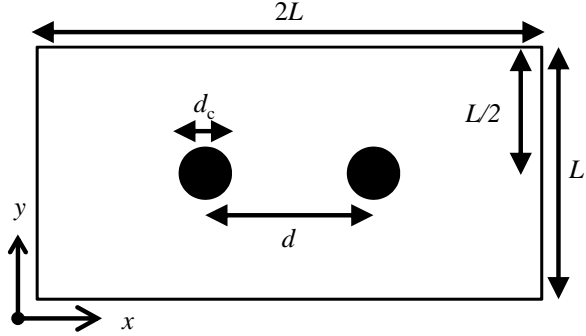}
\end{center}
\caption{Schematics of the system. 
Two infinitely long cylinders are placed in a rectangular system
with an area $L\times 2L$.
The length $L=30a_0$ 
and the diameter $d_{\rm c}=6a_0$ of both cylinders 
are fixed 
where $a_0$ is the unit length of the macroscopic fluid simulation.
The distance $d$ between two cylinders is a constant parameter.
The $x$-axis is chosen to be parallel to the flow direction 
and the $y$-axis is vertical to the $x$ axis.
Because of the translational symmetry towards 
the $z$-direction perpendicular to the $x$ and $y$ axes,
we can focus on the two dimmensional flow at the macroscopic level. 
When $d=L$, this system corresponds to 
the one considered in Ref.~\cite{MT2011}.
}
\label{fig.system}
\end{figure}

In Ref.~\cite{MT2011}, 
we have considered a flow around a cylindrical obstacle using the new method
and found that the polymer states are different 
between the upstream region and the downstream region
because of the flow-history-dependent behavior of 
the entangled polymer melt.
In that work, we considered the case that there is only one cylinder in
the system.
If there are two cylinders in the system as shown in Fig.~\ref{fig.system},
the flow history in the upstream region 
will affect the polymer states in the downstream region.
Moreover, we expect that the distance $d$ between two cylinders can affect 
the flow behavior of entangled polymer melt. 
Because 
the cylinders are placed in tandem along the flow direction,
the effect of the memory of the flow history can depend 
on the distance $d$ between the cylinders.
The purpose of the present paper
is to clearly show the flow-history-dependent behavior of entangled
polymer melt and 
to investigate the effect of the memory of the flow history
using the new multiscale simulation~\cite{MT2011}.

\section{Multiscale Simulation}

In the multiscale simulation employed here,
the polymer melt is described 
with an amount of fluid particles.
The position $\boldsymbol{r}_i$ and
the velocity $\boldsymbol{v}_i$ of
the $i$-th fluid particle
are updated according to
the following equations:
\begin{align}
\frac{{\rm d} \boldsymbol{r}_i}{{\rm d}t}&=\boldsymbol{v}_i,\\
\rho_i\frac{{\rm d} \boldsymbol{v}_i}{{\rm d}t}
&= -\boldsymbol{\nabla}p_i
+\boldsymbol{\nabla}\cdot \boldsymbol{\sigma}_i
+ \boldsymbol{f},
\end{align}
where $\boldsymbol{f}$ is 
an external body force with $\boldsymbol{f}=(f_x,0)$.
These equations are integrated 
with the velocity-Verlet algorithm.
The spatial derivative of a field variable $\boldsymbol{\nabla}f$ 
is calculated with the modified smoothed particle hydrodynamics 
(MSPH) algorithm\cite{MSPH,FPM}.
The density $\rho_i$ at $i$-th fluid particle is calculated with 
the usual smoothed particle hydrodynamics (SPH)~\cite{SPH} technique:
$\rho_i=\sum_{j\in\Omega_i}m_0 W(|\boldsymbol{r}_i-\boldsymbol{r}_j|,h)$
where $W(r,h)$ is a Gaussian function with a cutoff length $2h$\cite{MSPH}.
The pressure $p_i$ is a function of $\rho_i$.
The stress tensor $\boldsymbol{\sigma}_i$ 
depends on the microscopic states of polymers 
simulated with the polymer dynamics simulation.
Using the dual-slip-link model (DSLM)~\cite{DSLM_PRL,DSLM}
 as the microscopic simulation
in the multiscale simulation,
we can obtain the polymeric stress $\boldsymbol{\sigma}^{\rm p}$
coming from the entangled polymer states.
DSLM can describe the polymer dynamics 
larger than the tube diameter $b$.
To include the faster dynamics than the tube dynamics
(or the dynamics less than the tube diameter),
we supply the dissipative stress $\boldsymbol{\sigma}^{\rm d}$ 
to the stress tensor $\boldsymbol{\sigma}$.
Because the relaxation time of such faster dynamics
is negligibly small compared to the tube dynamics,
we can regard the dissipative stress $\boldsymbol{\sigma}^{\rm d}$
 as a Newtonian stress tensor
with a constant viscosity $\eta^{\rm d}$.
Then, the stress tensor $\boldsymbol{\sigma}_i$ 
is assumed to be
the sum of $\boldsymbol{\sigma}^{\rm p}$ and $\boldsymbol{\sigma}^{\rm d}$:
\begin{align}
\boldsymbol{\sigma}_i&=
\boldsymbol{\sigma}^{\rm p}_i
+\boldsymbol{\sigma}^{\rm d}_i.
\end{align}
The polymeric stress $\boldsymbol{\sigma}^{\rm p}_i$ is obtained 
from DSLM as follows:
\begin{align}
\boldsymbol{\sigma}^{\rm p}_i&=
\sigma_0
\left\langle\sum_{j=1}^{Z} 
\frac{\boldsymbol{r}_{j}^{\rm t} \boldsymbol{r}_{j}^{\rm t} }
{|\boldsymbol{r}_j^{\rm t}| b}
\right\rangle_i,\label{eq.sigma_p}
\end{align}
where $Z$ is the number of entanglements in a polymer chain,
$\boldsymbol{r}^{\rm t}_j$ is the $j$-th tube segment vector on the
chain,
and $\sigma_0\equiv (\eta^{\rm p}/\eta^{\rm p0})\sigma_{\rm e}$ 
is the stress coefficient that transforms 
the stress unit $\sigma_{\rm e}$
in the microscopic simulation to that in the macroscopic
simulation through the ratio $\eta^{\rm p}/\eta^{\rm p0}$.
The bracket $\langle \cdots \rangle_i$ means the average over 
the polymer chains in the $i$-th fluid particle.
The dissipative stress $\boldsymbol{\sigma}^{\rm d}$ is assumed 
to be proportional to the deformation rate tensor $\boldsymbol{D}$
as follows:
\begin{align}
\boldsymbol{\sigma}^{\rm d}_i&=\eta^{\rm d} \boldsymbol{D}_i,
\end{align}
where 
$\eta^{\rm d}$ is the dissipative viscosity
and
$\boldsymbol{D}_i\equiv 
\boldsymbol{\nabla} \boldsymbol{v}_i + 
(\boldsymbol{\nabla} \boldsymbol{v}_i)^{\rm T}$.

The macroscopic fluid particle simulation and the microscopic polymer
dynamics simulation communicate with each other 
through the following procedure:
\begin{enumerate}
\item $\boldsymbol{\kappa}_i
\equiv (\boldsymbol{\nabla}\boldsymbol{v}_i)$ 
is calculated in the fluid particle simulator
using MSPH algorithm.
\item $\boldsymbol{\kappa}_i$ is transfered to the polymer simulator.
\item The polymer states, namely $\{Z\}$ and $\{\boldsymbol{r}^{\rm t}_j\}$,
subjected to $\boldsymbol{\kappa}_i$ are updated for $\Delta t$
in the polymer simulator.
\item $\boldsymbol{\sigma}^{\rm p}_i$ is calculated using
      Eq.~\ref{eq.sigma_p}
\item $\boldsymbol{\sigma}^{\rm p}_i$ is transfered to the fluid
      particle simulator.
\end{enumerate}
The process (3) is composed of several steps, 
shortly explained as follows (for more information, see
Ref. \cite{DSLM}):
\begin{itemize}
\item[]
\begin{enumerate}
\item[(3.1)] Subjected to $\boldsymbol{\kappa}_i$, the tube segments 
$\{\boldsymbol{r}^{\rm t}_j\}$ are affinely deformed.
\item[(3.2)] The updated chain length 
$l=\sum^Z_{j=1}|\boldsymbol{r}_j^{\rm t}| + s_1 + s_2$ 
is relaxed only for a time interval $\Delta t$ where $s_1$ and $s_2$ are 
the lengths of end-segments 
of the polymer chain out of the tube;
The relaxation dynamics is composed of the reptation motion and 
the chain retraction where 
the equilibrium length of the polymer chains, 
$l_{\rm eq}\equiv \langle Z \rangle_{\rm eq}b$,
is assumed.
\item[(3.3)] The number of entanglement $Z$ is renewed 
depending on the length of $s_i$ $(i=1,2)$.
When $s_i$ is larger than $b$, a new tube segment 
with a length $b$ is created 
at the end of the chain in $s_i$ side, 
and $b$ is subtracted from $s_i$.
At the same time, a new entanglement point is created
in the other polymer chain randomly selected.
This new entanglement point and the new end of the polymer tube are
coupled with each other, which mimics the entanglement 
between the polymer chains.
When $s_i$ is less than $0$, the entanglement 
at the end of the polymer tube is erased 
and the entangled pair is also erased, 
which mimics the disentanglement process (constraint release).
\end{enumerate}
\end{itemize}
The procedure from (1) to (5) is the most time-consuming part
in the multiscale simulation
due to the excessively large number of degrees of freedom 
coming from the large number of fluid particles each having 
the large number of internal degrees of freedom.
However, 
because the procedure from (1) to (5) 
is independent of the other fluid particles,
parallel computing drastically improves the time-efficiency
of the multiscale simulation.

The cylinder is represented with the $N_{\rm c}$ fluid particles evenly 
placed on the perimeter.
These fluid particles fixed on the space
also have their own polymer simulator that guarantees 
the non-slip boundary condition assumed 
on the interface between the polymer melt and the wall~\cite{comment.1}.

The parameters used in the present work 
are summarized in Table~\ref{t.1}.
Using these parameters, we have found that the Reynolds number
${\rm Re}=\rho U d_{\rm c}/\eta_0$ is about 0.3 
and the Deborah number ${\rm De}=U\tau_{\rm d}/d_{\rm c}$ is about 1.8
where $U$ is the average flow velocity in the system and 
is set to $0.055a_0/t_0$ when $d=30a_0$ 
(cf., Ref.~\cite{MT2011}).

\begin{table}[t]
\caption{Fixed parameters in this article (cf., Ref.~\cite{MT2011}).}
\label{t.1}
\begin{tabular}{llll}
\hline
$t_0$ & &1.0 & Unit time in the macroscopic simulation\\
$a_0$ & &1.0 & Unit length in the macroscopic simulation\\
$m_0$ & &1.0 & Unit mass in the macroscopic simulation\\
$t_{\rm e}$ & $(=t_0)$ & 1.0 & Unit time in the microscopic simulation\\
$a_{\rm e}$ & $(=b \ll a_0)$ & 1.0 & 
Unit length in the microscopic simulation (Tube diameter)\\
$\sigma_{\rm e}$ & & $1.0$ & Unit stress in the microscopic simulation\\
$\Delta t$ & & $1.0\times 10^{-2}t_0$ & Time-integral step\\
$\langle Z \rangle_{\rm eq}$ & & $7.0$ & 
Average number of entanglements in equilibrium\\
$\eta^{\rm p0}$ & & $17.5 \sigma_{\rm e}t_{\rm e}$
& Zero-shear viscosity of entangled polymer melt 
when $\langle Z \rangle_{\rm eq}=7$~\cite{MT2011}\\
$\tau_{\rm d}$ & & $200t_{\rm e}$
& Disentanglement time (Reptation time) 
when $\langle Z \rangle_{\rm eq}=7$~\cite{MT2011}\\
$\eta_0$  & $(=\eta^{\rm p}+\eta^{\rm d})$ & $1.0 m_0/(a_0 t_0)$ &
Total zero-shear viscosity of the macroscopic flow\\
$\eta^{\rm p}/\eta_0$ &  & $0.9$ & Viscosity ratio of polymeric viscosity\\
$\eta^{\rm d}/\eta_0$ & & $0.1$ & Viscosity ratio of dissipative viscosity\\
$f_x$ & & $5.0\times 10^{-4} m_0/(a_0^2 t_0^2)$ & External body force\\
$L$ & & $30 a_0$ & System length\\
$d_{\rm c}$ & & $6.0 a_0$ & Diameter of the cylinders\\
$N_{\rm f}$ & & $1,784$ & Total number of fluid particles\\
$N_{\rm p}$ & & $1,000$ & Number of polymer chains in a fluid particle\\
$N_{\rm c}$ & & $24$ & Number of fluid particles that represent 
the perimeter of a cylinder\\
\hline
\end{tabular}
\end{table}

\section{Results and Discussions}

Now we investigate
entangled polymer melt flows
in the system shown in Fig.~\ref{fig.system} 
using the multiscale simulation.
At first, we focus on two different cases (a) and (b)
where the distances between two cylinders are (a) $d=10a_0$ and
(b)
$d=20a_0$ (cf., Ref~\cite{MT2011} when $d=30a_0$).
As is the case in Ref.~\cite{MT2011}, 
we focus on the magnitudes 
of the velocity field $|\boldsymbol{v}|$,
the shear deformation rate field $|D_{xy}|$,
and the shear stress field normalized 
by the total zero-shear viscosity
$|\sigma_{xy}|/\eta_0$.
The fluid particle data are transformed 
into the data on the square lattice mesh
using the linear interpolation method.
Then, to decrease the noise of the data,
we take an average of the obtained data 
over the time interval from $2,000 t_0$ to $3,000 t_0$
in the steady state,
because the macroscopic states of flow have reached steady states 
before $1,000 t_0$
in this system.

\begin{figure}[t]
\begin{center}
\includegraphics[width=\columnwidth]{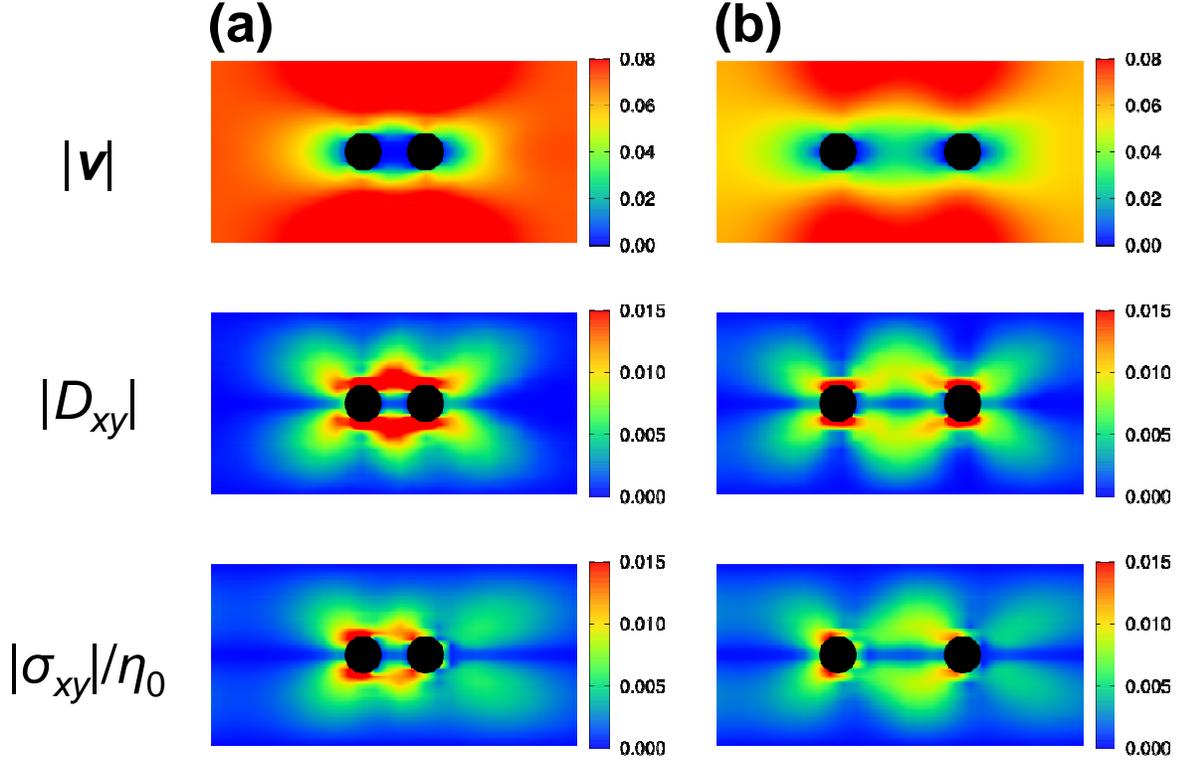}
\end{center}
\caption{Flow behaviors in cases of 
the distances (a) $d=10a$ and (b) $d=20a$ between two cylinders.
The magnitudes of the velocity $\boldsymbol{v}$,
the shear deformation rate $D_{xy}$ and 
the shear stress normalized by 
the total zero-shear viscosity $\sigma_{xy}/\eta_0$
are shown here.}
\label{fig.d10d20}
\end{figure}

The spatial distributions of these data are shown in Fig.~\ref{fig.d10d20}.
Comparing these two cases, as we expected,
the flow-history-dependent behavior is clearly appeared
in the shear stress distribution in (a) 
where the magnitude of the shear stress field around the cylinder in the
upstream side is clearly 
different from that around the other cylinder in the downstream side,
while it is difficult to observe such a behavior in (b).
Moreover, focusing on the velocity fields in (a) and (b),
the average flow velocity $U$ in (a) seems to be 
higher than that in (b) because the high velocity region 
in (a) is broader than that in (b).

To clarify the relationship between $U$ and $d$,
we plot the average flow velocity $U$ 
against the distance $d$ 
in Fig.~\ref{fig.ud} (a).
For the comparison, 
the open circles represent
the results in the Newtonian fluid
($\eta^{\rm d}=\eta_0$ and $\eta^{\rm p}=0$).
As shown in Fig.~\ref{fig.ud},
the average flow velocity $U$ decreases 
with increasing the distance $d$ when $d<30a_0$.
Because of the periodicity of the system,
the figure shows a symmetry on the line $d=30a_0$,
namely
the configuration of the cylinders with the distance $d$ for $d>30a_0$
corresponds to that for $|d-30a_0|$. 
Normalizing 
$U$ and $d$ 
as $U'=(U(d)-U(30a_0))/U(30a_0)$ and $d'=|d-30a_0|/30a_0$,
$U'$ shows a monotonic increasing behavior with $d'$ as shown 
in Fig.\label{fig.ud} (b).
The solid line 
and the dashed line in Fig.\label{fig.ud} (b) 
are fitting lines
obtained using the nonlinear least squares fitting.
The shapes of these fitting curves are 
$U'(d')= c_{\rm P} {d'}^{\alpha_{\rm P}}$ 
where $c_{\rm P}=1.411 \pm 0.016$ and $\alpha_{\rm P}=2.979 \pm 0.035$
for the polymer melt flow
and $U'(d')= c_{\rm N} {d'}^{\alpha_{\rm N}}$ 
where $c_{\rm N}=0.857 \pm 0.014$ 
and $\alpha_{\rm N}=2.969 \pm 0.051$ for the Newtonian flow.
Comparing these results,
the normalized flow velocity $U'$ of the polymer melt is
clearly enhanced with increasing $d'$,
and
the powers of the fitting functions for the polymer and Newtonian flows,
however, 
are found to 
have almost the same value; $c_{\rm P}>c_{\rm N}$ and
$\alpha_{\rm P}\simeq \alpha_{\rm N} \simeq 3$.
Namely,
the flow-history-dependent behavior enhances the average flow velocity
(or decreases the viscosity)
but does not affect the exponent of the power law function
in this flow system.

\begin{figure}[t]
\begin{center}
\includegraphics[width=\columnwidth]{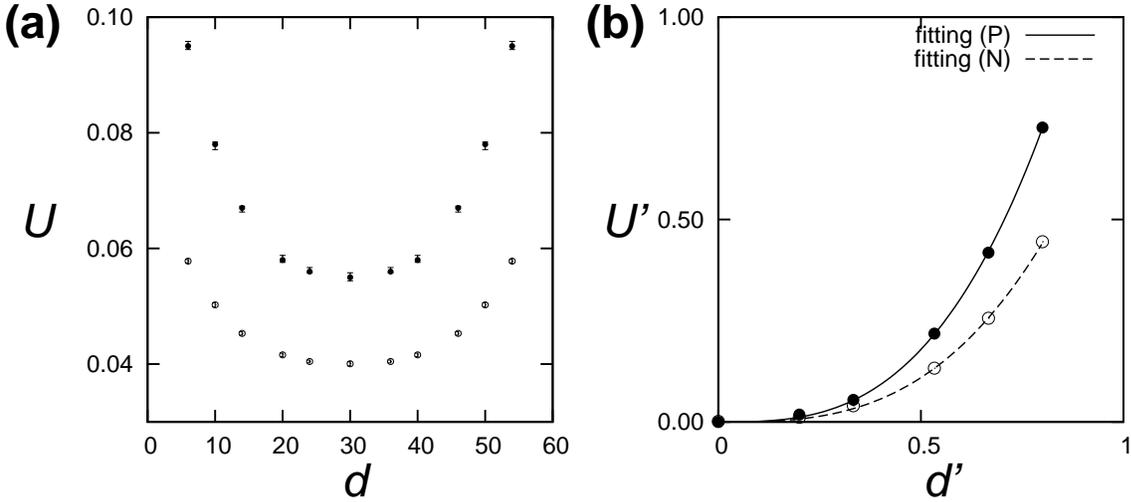}
\end{center}
\caption{The averaged flow velocity $U$ and the distance $d$ 
between two cylinders. 
The filled circles in the graph represent the polymer melt flow.
For the comparison, the Newtonian flow results
($\eta^{\rm d}=\eta_0, \eta^{\rm p}=0$) are also shown in the graph 
using open circles.
When $d$ is less than $30a$, 
the averaged flow velocity $U$ decreases 
with increasing the distance $d$.
Because of the periodicity of the system, 
this graph shows a symmetry on $d=30a_0$, namely
$U(d)$ for $d>30a_0$ corresponds to $U(60a_0-d)$.
Normalizing $U$ and $d$ as 
 $U'=(U(d)-U(30a_0))/U(30a_0)$
and $d'=|d-30a_0|/30a_0$,
the normalized velocity $U'$ is found to be proportional to $d'^{3}$
both in the polymer melt flow and the Newtonian flow.}
\label{fig.ud}
\end{figure}

\section{Summary}

We have investigated entangled polymer melt flow 
around two cylinders in tandem 
in a two dimensional rectangular system, 
using the new multiscale simulation\cite{MT2011}
that is composed of the macroscopic fluid particle simulation
and the microscopic entangled polymer dynamics simulation.
In the system, the polymer melt has represented
the flow-history-dependent behavior in the shear stress distribution. 
We have found that 
the flow-history-dependent behavior causes 
the orientation of polymer chains,
and then reduces the flow resistance.
However, the flow history does not affect
the exponent of the power law function obtained 
fitting the average flow velocity against the distance between cylinders.


\begin{thebibliography}{9}
\bibitem{comment.1}
So far, only non-slip boundary condition 
between the polymer melt and the wall
is manageable in the present multiscale simulation.
Substituting the bulk polymer simulator 
in the fluid particles fixed on the space
to a simulator 
accounting for an interfacial dynamics among the polymer melt 
and the wall,
we will be able to discuss the slip flow
in the multiscale simulation.
However, the slip behavior is observed in the very high Weissenberg
number flow where ${\rm Wi} \gg 1.0$, which far exceeds the flow ranges
discussed in this work.
\end{thebibliography}
\end{document}